\documentclass[%
 twocolumn,
 reprint,
 superscriptaddress,
showpacs,
 amsmath,amssymb,
 aps,
]{revtex4}
\usepackage{color}
\usepackage{ulem}
\usepackage{graphicx}
\usepackage{dcolumn}
\usepackage{bm}
\usepackage{here}
\usepackage{url}

\definecolor{myorange}{rgb}{0.7,0.5,0.0}
\definecolor{mygreen}{rgb}{0.0,0.7,0.0}

\definecolor{purple}{rgb}{0.54, 0.17, 0.89}

\begin{document}

\preprint{APS/123-QED\author{Takahiro Ohgoe}}

\title{ 
  {\it Ab Initio} Study on Superconductivity and Inhomogeneity in Hg-based Cuprate Superconductor}

\author{Takahiro Ohgoe}
\altaffiliation[present address: ] {Research Institute for Science and Engineering, Waseda University, Okubo, Shinjuku-ku, Tokyo 169-8555, Japan, \\
$^{\dagger}$ present address: Toyota Physical and Chemical Research Institute, Yokomichi, Nagakute, Aichi  480-1192, Japan}

\affiliation{%
  Department of Applied Physics, The University of Tokyo, Hongo, Bunkyo-ku, Tokyo, 113-8656, Japan
}%

\author{Motoaki Hirayama}
\affiliation{%
        RIKEN Center for Emergent Matter Science, Wako, Saitama 351-0198, Japan
}%
\author{Takahiro Misawa}
\affiliation{%
        Institute for Solid State Physics, The University of Tokyo, Kashiwanoha, Kashiwa, Chiba 277-8581, Japan
}%
\author{Kota Ido}
\affiliation{%
        Institute for Solid State Physics, The University of Tokyo, Kashiwanoha, Kashiwa, Chiba 277-8581, Japan
}%
\author{Youhei Yamaji}
\affiliation{%
        Department of Applied Physics, The University of Tokyo, Hongo, Bunkyo-ku, Tokyo, 113-8656, Japan
}%
\affiliation{%
        JST, PRESTO, Hongo, Bunkyo-ku, Tokyo, 113-8656, Japan
}%
\author{Masatoshi Imada$^{*,\dagger}$}%
\affiliation{%
        Department of Applied Physics, The University of Tokyo, Hongo, Bunkyo-ku, Tokyo, 113-8656, Japan  
}%
\date{\today}

\begin{abstract}
Understanding physics of high-$T_c$ cuprate superconductors remains one of the important problems in materials science. Though a number of diverse theories argue about the superconductivity and competing orders, {\it ab initio} and quantitative understanding is lacking.  Here, we reproduce the experimental phase diagram of HgBa$_2$CuO$_{4+y}$ by solving its {\it ab initio} low-energy effective Hamiltonian without adjustable parameters. It shows a superconducting phase in a wide range of hole density $\delta$, and its competition with charge period-4 plus spin period-8 stripe order near $\delta \sim 0.1$, in agreement with experimental results including recent X-ray scattering. Then a crucial role of off-site interactions in stabilizing the superconductivity is elucidated with emphasis on charge fluctuations. It also clarifies the condensation energy mainly contributed from the onsite Coulomb interaction. The present achievement will enable deeper, predictable understanding on open issues of the high-$T_c$ superconducting mechanism and promote {\it ab initio} studies on strongly correlated electrons beyond parametrized model studies.

\end{abstract}

\pacs{71.10.Fd, 71.27.+a, 74.72.-h}
\maketitle


\section{\label{sec:intro}Introduction}
Since the discovery of high-$T_c$ cuprates, enormous number of experimental reports have unveiled their rich and complex physics, which have shed light on mechanisms of superconductivity. Especially in the underdoped region, unconventional phenomena such as pseudogap, nematicity\cite{kivelson1998,hinkov2008,ysato2017} and stripe order\cite{tranquada1997,keimer2015} were observed and they are still intensively studied both experimentally and theoretically. Owing to recent advancement of experimental tools such as the scanning tunneling microscope (STM), resonant X-ray scattering, and X-ray diffraction imaging, charge orders (spatial inhomogeneity) have been widely reported in the underdoped region of several families of high-$T_c$ cuprates adjacent to superconducting phase, signaling their presence as a common feature\cite{wu2012,fink2009,chang2012,chiringhelli2012,vershinin2004,howald2003,wise2008,tabis2014,campi2015,comin2016,wu2011,wu2015}.

Historically, full theoretical understanding of the complex physics in high-$T_c$ cuprates has been hampered for long years, partly because previous theoretical approaches were mostly based on simple models with adjustable parameters and/or they are solved approximately at various levels. These limitations yielded diverse theoretical proposals which are often controversial with each other and relevance to real materials is not well established because of the uncertainty about adequacy of assumed parameters and the diversity in experimental indications. However, owing to the recent development of {\it ab initio} methods without relying on adjustable parameters and tools to solve them accurately, we are now at the stage of overcoming at least some of these controversies: Methods of deriving {\it ab initio} low-energy effective Hamiltonians, utilizing several tools such as the maximally localized Wannier function\cite{marzari1997,souza2001} and the constrained random phase approximations\cite{aryasetiawan2004}, were developed to construct a parameter-free theory. In the procedure to solve thus derived {\it ab initio} low-energy effective Hamiltonians, recent progress in accurate numerical methods has opened a possibility to finally reach conclusive results without adjustable parameters (see Appendix A for other attempts of {\it ab initio} studies).

In fact, on the level of model studies with parameters, carrier doped Hubbard model on a square lattice, one of the well-known simple models for the cuprates has been solved by state-of the-art numerical tools and its ground state has shown overall consensus indicating the dominance of charge inhomogeneous state such as charge and spin stripe state, severely competing with $d$-wave superconductivity in a wide range of doping concentration\cite{m_kato1990, white2000,capone2006,misawa2014,otsuki2014,corboz2016,tocchio2016,zhao2017,zheng2017,ido2018,darmawan2018}. However, neither the wavelength of the spin/charge order nor the wide region of the homogeneous superconducting ground state is quantitatively consistent with those observed in the cuprates\cite{tranquada1997,wu2012,fink2009,chang2012,chiringhelli2012,vershinin2004,howald2003,wise2008,tabis2014,campi2015,comin2016}. This shows the necessity of quantitative parameter-free studies to predict or reproduce the physics of real materials beyond the model study. Therefore, accurate first-principles studies of the microscopic Hamiltonian without adjustable parameters are desired to make an essential step forward to complete understanding of the long-standing issue on physics of the cuprate superconductors.

Here, we study an {\it ab initio} low-energy effective Hamiltonian derived for the high-$T_c$ cuprate HgBa$_2$CuO$_{4+y}$\cite{hirayama2018,hirayama2019} by using a many-variable variational Monte Carlo (mVMC) method\cite{tahara2008,misawa2019} and its refinement by combining with the fat-tree tensor network\cite{zhao2017} and/or the power Lanczos method\cite{heeb1993} together with variance extrapolations of energies to reach sufficient accuracy. We found that a quantitative evaluation of effects from off-site Coulomb interactions is crucial to reproduce $d$-wave superconductivity stabilized against the charge order as observed in the experimental results on the cuprates, in contrast to the charge-order dominance found in the simple Hubbard model. To our knowledge, this is a first-ever quantitative reproduction of the dominance of superconductivity in the cuprates without any adjustable parameters despite the severe realistic competition with the charge inhomogeneities. Such a quantitative reproduction is an important and imperative step for further understanding on the mechanism and future design for better functionality. We then elucidate a strong positive correlation between the enhancement of superconductivity and that of uniform charge susceptibility.

In Sec. \ref{sec:ham}, we describe the {\it ab initio} low-energy effective Hamiltonians which we will analyze. The detail of our numerical method is explained in Sec. \ref{sec:method}. The results for homogeneous states are shown in Sec. \ref{sec:homo}. Then, we show the results for inhomogeneous states in the subsequent Sec. \ref{sec:inhomo}. In Sec. \ref{sec:offsite_coulomb}, we analyze effects of off-site screened Coulomb interactions. We also present the results which analyze the connection between charge fluctuations and superconductivity in Sec. \ref{sec:chi_c}. Finally, we discuss and summarize our results in Sec. \ref{sec:sum}.

\section{\label{sec:ham} {\it Ab initio}  effective Hamiltonians}
In a previous work, Hirayama {\it et al.} derived low-energy effective Hamiltonians for ${\rm Hg Ba}_2 {\rm Cu O}_4$ and ${\rm La}_2{\rm Cu O}_4$ from first principles\cite{hirayama2018,hirayama2019}. In this derivation, they employed the constrained GW calculations supplemented by the self-interaction correction (cGW-SIC) to remove the double counting of the exchange correlations\cite{hirayama2013,hirayama2017}. To derive the screened Coulomb interactions, the constraint random phase approximation was employed\cite{aryasetiawan2004}.
   Reference \cite{hirayama2019} further employed the procedure of the self-consistent feedback of interband interaction between the low-energy and high-energy degrees of freedom by considering the pinning of orbital occupation by following the spirit studied before \cite{pourovskii2007}. The feedback treatment in Ref. \cite{hirayama2019} is the following: When the effective cGW Hamiltonian is solved, the obtained low-energy orbital occupation may differ from the GW charge distribution in general. However, the electrons contained in a large number of bands outside the degrees of freedom of the effective Hamiltonian impose strong (Hartree) potential, which generates the constraint to pin the orbital occupation rather than on the chemical potential for the electrons in the low-energy degrees of freedom. Therefore, each orbital filling should be preserved when one solves the effective Hamiltonian\cite{pourovskii2007,bhandary2016}.
    In this study, we employ the {\it ab initio} single-band effective Hamiltonian for the target antibonding orbital of ${\rm Hg Ba}_2 {\rm Cu O}_4$ thus derived in Ref. \cite{hirayama2019}, which takes the form of 
\begin{eqnarray}
  {\cal H}  =  -  \sum_{\sigma} \sum_{i,j} t_{ij} c_{i \sigma}^{\dagger}c_{j \sigma} +  \sum_i U n_{i \uparrow} n_{i \downarrow} +  \sum_{i<j} V_{ij} n_i n_j. 
\label{Hamiltonian}
\end{eqnarray}

We consider the two-dimensional ${\rm CuO}_2$ plane with $i$, $j$ representing unit cell indices, where the maximally localized Wannier function is constructed for the molecular orbital\cite{marzari1997,souza2001}. $c_{i \sigma}^{\dagger}$ ($c_{i \sigma}$) is the creation (annihilation) operator of electrons with spin $\sigma$ (=$\uparrow$ or $\downarrow$) at the $i$-th Wannier orbital, and the number operator is $n_i = \sum_{\sigma} n_{i\sigma}$ with $n_{i\sigma}=c^{\dagger}_{i\sigma} c_{i\sigma}$. Here, $t_{ij}$ is the hopping parameters depending on the relative coordinate vector $\bm{r}_i -\bm{r}_j$, where $\bm{r}_i$ is the position vector of the center of the $i$-th Wannier orbital. $U$ and $V_{ij}$ are the screened on-site and off-site Coulomb interactions, respectively. Dominant component of the {\it ab initio} values derived in Ref. \cite{hirayama2019} are quoted here in Table \ref{tab:abval} for the self-contained description. The derived screened Coulomb interaction still decays as $\sim 1/r$ because the metallic screening is excluded in the derivation of the {\it ab initio} low-energy effective Hamiltonian. Therefore, we employ the Ewald summation method to treat its long-range part accurately without truncation\cite{ewald1921} (see Appendix B). On the other hand, the hopping parameters are short-ranged and it is enough to include them up to the third-nearest-neighbor hopping. We note that the off-diagonal interaction parameters other than the density-density interactions are small ($< 0.015U$), and thus can be ignored. In this work, we analyze the above Hamiltonian on square lattices with $N=L\times L$ sites. 
When hole carriers are doped into the Mott insulator at half filling $\langle n \rangle=\sum_{i\sigma} \langle n_{i\sigma} \rangle/N=1$, several different states are severely competing, and therefore highly accurate wavefunctions are required to determine the ground states.

\begin{table}[H]
\begin{center}
\scalebox{1.25}{
\footnotesize
\begin{tabular}{lccccc}
\hline \hline
 One-body   &  $t_1$ & $t_2$ & $t_3$ & $t_4$ & $t_5$  \rule{0pt}{2.6ex} \\
 parameters (eV)  & 0.509 & -0.127 & 0.077 & -0.018 & -0.004\rule{0pt}{2.6ex} \\
 Two-body  &  $U$ & $V_1$ & $V_2$ & $V_3$ & $V_4$ \rule{0pt}{2.6ex} \\
 parameters (eV) & 3.846 & 0.834 & 0.460 & 0.318 & 0.271  \rule{0pt}{2.6ex} \rule[-1.2ex]{0pt}{0pt} \\
\hline \hline
\end{tabular}
}
\caption{Derived parameters of the {\it ab initio} effective Hamiltonian. {\it Ab initio} hopping amplitudes and screened Coulomb interactions derived for the single-band effective Hamiltonian for ${\rm Hg Ba}_2 {\rm Cu O}_4$\cite{hirayama2019}. $t_n$ and $V_n$ represent the $n$-th nearest-neighbor hopping parameters and Coulomb interactions, respectively. The long-range part of off-site Coulomb interactions is obtained by fitting the available data to the $1/r$ function. (see Appendix B). Other off-diagonal Coulomb interactions are negligible.}
\label{tab:abval}
\end{center}
\end{table}

\section{\label{sec:method}Numerical method}
In our simulations, we used the many-variable variational Monte Carlo method\cite{misawa2014,tahara2008,misawa2019}. Our variational wave function takes the following form: $| \psi \rangle = {\cal P}^{\rm G} {\cal P}^{\rm J} {\cal P}^{\rm d-h}  | \phi^{\rm pair} \rangle$. Here, ${\cal P}^{\rm G} = \exp \left( \sum_{i} \alpha_i^{\rm G} n_{i \uparrow} n_{i \downarrow} \right)$, ${\cal P}^{\rm J} = \exp \left( \sum_{i<j} \alpha_{ij}^{\rm J} n_{i} n_{j} \right)$ and ${\cal P}^{\rm d-h} = \exp \left[ - \sum_{m=0}^4 \sum_{l=1,2} \alpha_{m}^{(l)} \sum_{i} \xi_{i(m)}^{(l)} \right]$ are the Gutzwiller factor\cite{gutzwiller1963}, the long-range Jastrow correlation factors\cite{jastrow1955, capello2005}, and the doublon-holon correlation factor\cite{yokoyama1990}, respectively. $\xi_{i(m)}^{(l)}$ is the diagonal operator in the real-space representations which takes unity when a doublon (holon) exists at the $i$th site and $m$ holons (doublons) exist at the $l$th nearest neighbor. Otherwise, it takes zero. $\alpha$'s are the coefficients which should be optimized. In practice, we impose the translational symmetry on them. $| \phi^{\rm pair} \rangle$ is the generalized pairing wave function defined by $| \phi^{\rm pair} \rangle = \left( \sum_{i \sigma,j \sigma'} f_{i \sigma, j \sigma'} c_{i \sigma}^{\dagger} c_{j \sigma'}^{\dagger}  \right)^{N_{\rm e}/2} | 0 \rangle$, where $f_{i \sigma, j \sigma'}$ are variational parameters and $N_{\rm e}$ is the total number of electrons. We usually consider the case of $\sigma=\uparrow$ and $\sigma'=\downarrow$. This can be regarded as a generalization of the Hartree-Fock-Bogoliubov type wave function with AF/CO and SC orders\cite{tahara2008,giamarchi1991}, and thus flexibly describes these states as well as paramagnetic metals. In order to reduce the number of independent variational parameters, we assume that $f_{ij}$ have a sublattice structure such that $f_{ij}$ depend on the relative vector ${\bm r}_i-{\bm r}_j$ and a sublattice index of the site $j$ which we denote as $\eta(j)$. Thus, we can rewrite it as $f_{\eta(j)}({\bm r}_i-{\bm r}_j)$. In the present study on the homogeneous states, we assumed a 2$\times$2 sublattice structure. In this case, the number of independent $f_{ij}$ reduces from $N^2$ to $2 \times 2 \times N$. For studies on the C$l_c$S$l_s$ stripe states, we extended the sublattice structure of $f_{ij}$ to $l_s \times 2$, where $l_c$ is a fraction of $l_s$. We consider systems under the periodic-antiperiodic boundary condition.

In doped regions, the superconducting state and stripe states as well as the antiferromagnetic state are severely competing. To determine the lowest energy state among them, highly accurate results of energies are required. Therefore, we performed extrapolations of energies to the zero-variance limit\cite{kwon1993,imada2000, sorella2001}. For this purpose, we obtained improved energies by combining the fat-tree tensor network\cite{zhao2017} and/or performing the 1st Lanczos step. In recent studies, it has been shown\cite{darmawan2018} that for the simple Hubbard model, the energies obtained by the same procedure have the same level of accuracy with those obtained by the different state-of-numerical methods\cite{zheng2017}. Examples of the extrapolations in the present studies are shown in Fig. \ref{fig:var_extrapolation} (a) of the Appendix E ($L$=24). In Fig. \ref{fig:var_extrapolation} (b) and (c), we also present the results for different system sizes ($L$=18, 24, 30) to show the size dependence of the extrapolated energies.

\section{\label{sec:homo}Homogeneous states}

\begin{figure}[h]
\begin{center}
  \includegraphics[width=9cm]{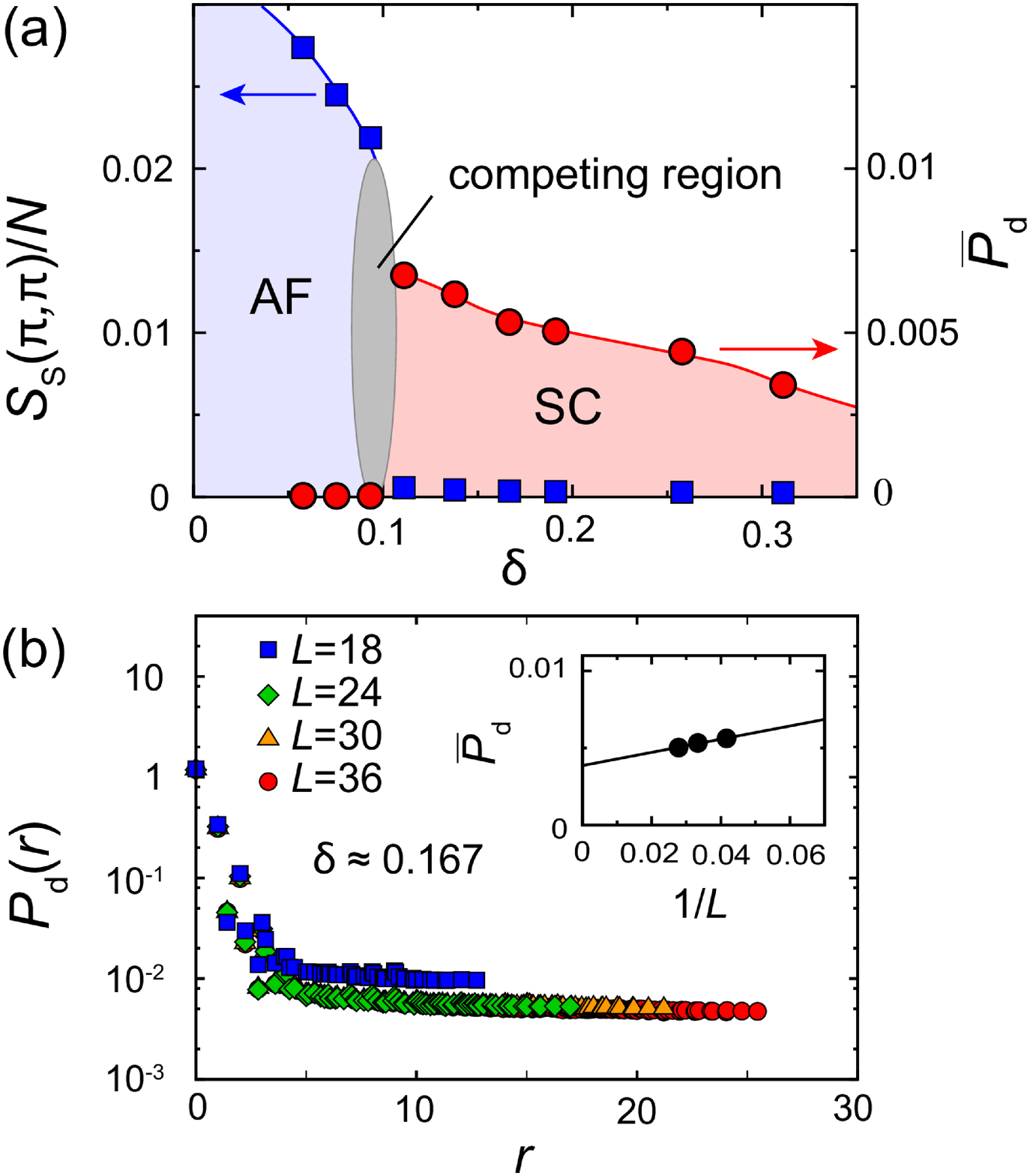}
\caption{(Color online)  (a) Physical quantities $S_s(\pi, \pi)/N$ and ${\overline P}_{d}$ of homogeneous states ($L=30$) as functions of $\delta$. Gray region indicates a region where AF, SC, and a stripe state are severely competing. [For the energy competition with stripe states, see Fig. \ref{fig:stripes}]. (b) Size dependence of $P_{d}(r)$ at $\delta \simeq 0.167$. In the inset, ${\overline P}_{d}$ ($L=24$, 30, and 36) is extrapolated to the thermodynamic limit.}\label{fig:phys_uniform}\end{center}
\end{figure}

We first study charge-homogeneous states. Here, we assumed the $2\times 2$ sublattice structure for our variational wave function\cite{misawa2019}. The measured physical quantities are the spin structure factor $S_s({\bm q}) = \frac{1}{3 N} \sum_{i,j} \langle {\bm S}_i \cdot {\bm S}_j \rangle e^{i {\bm q}\cdot  ({\bm r}_i-{\bm r}_j)}$ (${\bm S}_i$  is the spin operator at the site $i$) and the simple average of the $d$-wave superconducting correlation function over the long-range part: ${\overline P}_{d} = \frac{1}{M} \sum_{\sqrt{2}L/4 < r} P_{d}({\bm r})$, where ${\bm r}$ belongs to $(-L/2, L/2]^2$ and $M$ is the number of lattice points satisfying $\sqrt{2}L/4<r=|{\bm r}|<\sqrt{2}L/2$. The correlation function is defined by $ P_{d}({\bm r}) = \frac{1}{2N} \sum_{{\bm r}_i} \langle \Delta_{d}^{\dagger} ({\bm r}_i) \Delta_{d} ({\bm r}_i+{\bm r}) + \Delta_{d} ({\bm r}_i) \Delta_{d}^{\dagger} ({\bm r}_i+{\bm r}) \rangle$ with the order parameter $\Delta_{d}({\bm r}_i) = \frac{1}{\sqrt{2}} f_{d}({\bm r}) (c_{{\bm r}_i \uparrow} c_{{\bm r}_i + {\bm r}\downarrow} - c_{{\bm r}_i \downarrow} c_{{\bm r}_i + {\bm r}\uparrow} )$. $f_d({\bm r})$ is the ${d_{x^2-y^2}}$ form factor defined by $f_d({\bm r}) = \delta_{r_y, 0} (\delta_{r_x,1} + \delta_{r_x,-1}) - \delta_{r_x.0} (\delta_{r_y, 1} + \delta_{r_y, -1})$.
In Fig. \ref{fig:phys_uniform} (a), we plot $S_s(\pi, \pi)/N$ and ${\overline P}_{d}$  as functions of the doping concentration $\delta =1-\langle n \rangle$ at $L=30$. Here, we find two phases: antiferromagnetic (AF) phase for $\delta \lesssim 0.1$ and superconducting (SC) phase becomes the ground state for $\delta \gtrsim 0.1$. Typical size and spatial dependences of $P_{d}({\bm r})$ for the SC ground state are shown at $\delta \simeq 0.167$ in Fig. \ref{fig:phys_uniform} (b). The ground-state phase diagram shown in Fig. \ref{fig:phys_uniform} indicates that the $d$-wave superconducting state is the ground state in an extended region of doping concentration in the thermodynamic limit in agreement with the experimental phase diagram.

 Around $\delta \simeq 0.1$, the physical properties in Fig. \ref{fig:phys_uniform} (a) sharply change, which is indicative of a first-order transition. However, in the presence of long-range Coulomb interactions, the macroscopic phase separation is forbidden, and instead, it is replaced by other phases such as stripes or mesoscopic mixture of two competing phases (micro-emulsions)\cite{spivak2004,jamei2005,emery1993}. Indeed, we will show in the subsequent paragraphs that a stripe state intervenes in this region.

\begin{figure}[h]
\begin{center}
  \includegraphics[width=7cm]{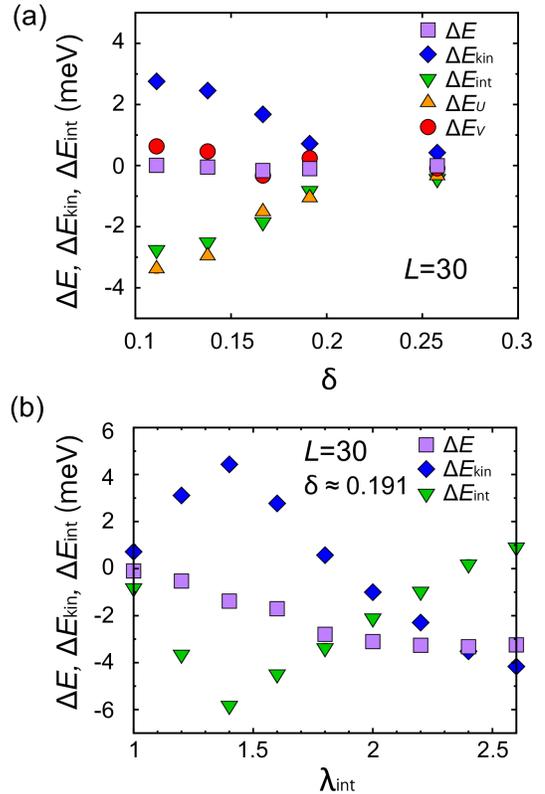}
\caption{(Color online) Super-Normal energy difference. (a) Super-Normal energy difference ($=-E_{\rm cond}$) $\Delta E, \Delta E_{\rm kin}, \Delta E_{\rm int}, \Delta E_{U}$ and $\Delta E_{V}$ as functions of $\delta$. (b) $\lambda_{\rm int}$-dependence of the super-normal energy difference. Since, for large $\lambda_{\rm int}$ beyond the realistic value $\lambda_{\rm int}=1$, the antiferromagnetic order often develops during the optimization process of the SC or normal state, we imposed the translational symmetry on $| \phi^{\rm pair} \rangle$ to exclude the antiferromagnetism and discuss the condensation energy between pure SC and normal states.}\label{fig:e_cond}\end{center}
\end{figure}

A question arises regarding the character of the observed SC state: Whether the SC state is interaction-energy driven or kinetic-energy driven in the {\it ab initio} Hamiltonian. In VMC studies\cite{misawa2014,yokoyama2013,tocchio2016} and cluster dynamical mean-field theory (cDMFT) studies\cite{gull2012,fratino2016} on the Hubbard model, it was observed that the character changes from interaction-energy driven to kinetic-energy driven at some intermediate values of $U/t_1$, with $t_1$ being the nearest-neighbor hopping, although there is quantitative differences in its values. However, studies on {\it ab initio} Hamiltonians to see which is correct in reality are missing. To examine it using our {\it ab initio} Hamiltonian, we calculated the energy difference between SC and normal (paramagnetic) states: $\Delta E = E_{\rm SC}/N - E_{\rm Normal}/N$, $\Delta E_{\rm kin} = E_{\rm kin, SC}/N - E_{\rm kin, Normal}/N$, and $\Delta E_{\rm int} = E_{\rm int, SC}/N - E_{\rm int, Normal}/N$. The subscripts ``kin'' and ``int'' denote the kinetic part [the first term of Eq. (1)] and the interaction part (the sum of other (second and third) terms) of energies, respectively. Note that the conventional definition of the condensation energy $E_{\rm cond}=-\Delta E$ has the opposite sign, where $\Delta E$ is negative when the superconducting state has lower energy. The results are obtained by the mVMC method (without variance extrapolations) and are shown on the doing concentration dependence at {\it ab initio} parameters in Fig. \ref{fig:e_cond} (a). Here, we also plotted the contributions from the on-site interaction part $\Delta E_{U}$ and the off-site interaction part $\Delta E_{V}$ of $\Delta E_{\rm int}$ (i.e. $\Delta E_{\rm int}=\Delta E_{U}+\Delta E_{V}$) separately in the plot. From Fig. \ref{fig:e_cond} (a), we observe that the SC state is decisively interaction-energy driven at {\it ab initio} parameters ($U/t_1 \sim 7.56$). The main contribution of the gain of the condensation energy $(-\Delta E)$ is clearly from the on-site interaction part. This indicates that the main source of the energy gain of the superconducting state is attributed to the reduced energy cost of the double occupation in the superconducting state. This is because the double occupation is prohibited by symmetry for the $d$-wave pair. Although there exists an uncertainty in the decomposition into the interaction and the kinetic energy parts depending on the choice of the number of electronic orbitals considered\cite{norman2000}, the present conclusion about the interaction driven superconductivity is unambiguous and firm for the {\it ab initio} single-band effective Hamiltonian.

We remark that the energy of the SC state is also severely competing with the paramagnetic normal metal, in contrast to more stable SC state found in the Hubbard model\cite{misawa2014}. The interpolation between the {\it ab initio} effective Hamiltonian and the Hubbard model in the strong coupling region reveals that the stable SC states in the Hubbard limit, which is well separated from the non-SC excited state, is adiabatically connected to the SC state in the {\it ab initio} limit, which is highly degenerate with the normal metal within the accuracy of the present method (see Appendix C). These nearly degenerate states are consistent with experiments since the experimentally estimated condensation energy is as small as 0.1 meV\cite{billon1997,kirtley1998}, which is beyond any available numerical method including the present numerical accuracy ($\sim 1-2$ meV).

\section{\label{sec:inhomo}Inhomogeneous states}
\begin{figure}[h!]
\begin{center}
  \includegraphics[width=8.5cm]{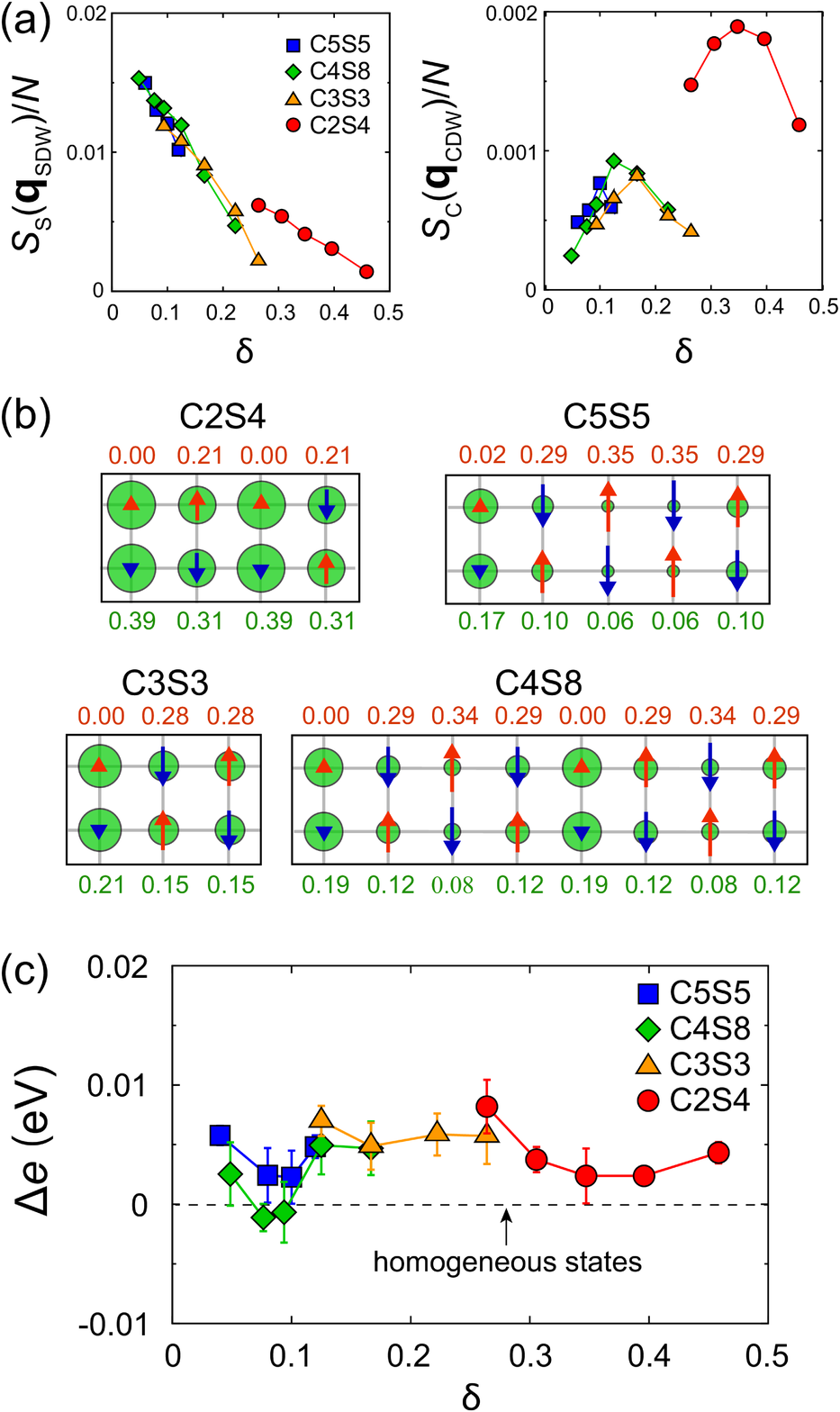}
\caption{(Color online) Physical quantities of inhomogeneous states. (a) $S_s({\bm q}_{\rm SDW})/N$ and $S_c({\bm q}_{\rm CDW})/N$ of stripe states as functions of $\delta$. ${\bm q}_{\rm SDW}$ and ${\bm q}_{\rm CDW}$ are the momenta at the peak of $S_s$ and $S_c$, respectively. The linear system sizes are $L=20$ for C5S5 and $L=24$ for others. The size dependence is small except for C2S4. For detailed size dependence, see Appendix D. (b) Spin/charge configurations of several stripes. The hole concentrations are $\delta=0.1$ for C5S5, $\delta=0.125$ for C4S8, $\delta \simeq 0.167$ for C3S3 and $\delta \simeq -.347$ for C2S4. In the same way as the previous studies (13), we represent the hole density $\delta = 1 - \langle n_i \rangle $ and the local spin moments $\langle n_{i,\uparrow} -n_{i,\downarrow} \rangle/2$ by the circle radius and the arrow length, respectively. Their values are also given as the green numbers and red numbers, respectively. (c) Stripe state energies relative to homogeneous states.}\label{fig:stripes}\end{center}
\end{figure}

We next consider charge inhomogeneous states. To describe states with long-period structures such as stripe states, we employ larger sublattice sizes imposed on the pair-product part of the variational wave function. In Fig. \ref{fig:stripes} (a), we present physical quantities of stripe states which are competing with homogeneous states. Here, the charge structure factor $S_c({\bm q}) = \frac{1}{N} \sum_{i,j} \langle n_i n_j \rangle e^{i {\bm q} \cdot ({\bm r}_i-{\bm r}_j)}$ is plotted as well as $S_s({\bm q})$. The wave vector ${\bm q}$ at the peak of the structure factors is described as $q_{\rm SDW}$ or  $q_{\rm CDW}$. ``C$l_c$S$l_s$'' represents charge/spin stripes with the period of $l_c$/$l_s$ in one direction parallel to the nearest neighbor Cu-Cu bond, whereas in the vertical direction, there are only antiferromagnetic spin modulations with the wavelength of 2 unit cells. The real-space spin/charge configurations are shown for its unit cell of symmetry broken state in Fig. \ref{fig:stripes} (b). Since the energies of stripes with $l_c \ge 6$ are higher than those with $l_c \le 5$, we do not include them here. The spin and charge structure factors divided by the system size show that the spin orders are monotonically decreasing as $\delta$ increases, and the charge orders have dome structures\cite{ido2018}, whereas superconducting correlation ${\overline P}_{d}$ is extrapolated to vanishingly small values in the charge inhomogeneous state (see Appendix D).

To clarify the energy differences, we show the energies of stripe states relative to the homogeneous states in Fig. \ref{fig:stripes} (c). Experimentally, the wave vectors of charge orders observed in the underdoped region of the hole-doped high-$T_c$ cuprates are $q \sim 0.15 - 0.35$ r.l.u. (reciprocal lattice unit) in the $a$-axis\cite{tranquada1997,wu2012,fink2009,chang2012,chiringhelli2012,vershinin2004,howald2003,wise2008,tabis2014,campi2015,comin2016}. In our results, the stripes with $l_c=3-5$ corresponding to wave vectors $q \sim 0.1 - 0.33$ r.l.u. are competing with homogeneous states in the underdoped region $\delta<0.15$. However, the stripe states have lower (or at least very close) energies only around $\delta \sim 0.1$.  This should be contrasted with the stripe ground state dominating a wide doping concentration for the simple Hubbard model\cite{zheng2017,ido2018,darmawan2018} and shows the importance of using {\it ab initio} values for the Hamiltonian parameters to describe the competition in real materials. For ${\rm Hg Ba}_2 {\rm Cu O}_{4+y}$ recent X-ray scattering experiments observed charge orders with $q \simeq 0.23$ 
for $\delta \simeq 0.12$\cite{campi2015} and $q \simeq 0.28$ for $\delta \simeq 0.09$\cite{tabis2014}. In our results of Fig. \ref{fig:stripes} (c), the stripe with $l_c=4$ ($q = 0.25$) is particularly competitive for $\delta \sim 0.1$, which is close to the experimental observations. This is again different from the stripe period of $l_c>5$ stabilized in the simple Hubbard model for $\delta \sim 0.1$\cite{darmawan2018}. Our extrapolation of the charge orders indicates that they have small but nonzero values in the thermodynamic limit (Appendix D), whereas the experimentally observed charge orders are short-ranged, probably partly due to disorder or impurity effects.

\section{\label{sec:offsite_coulomb}Effects of off-site screened Coulomb interactions}
\begin{figure}[h]
\begin{center}
  \includegraphics[width=8cm]{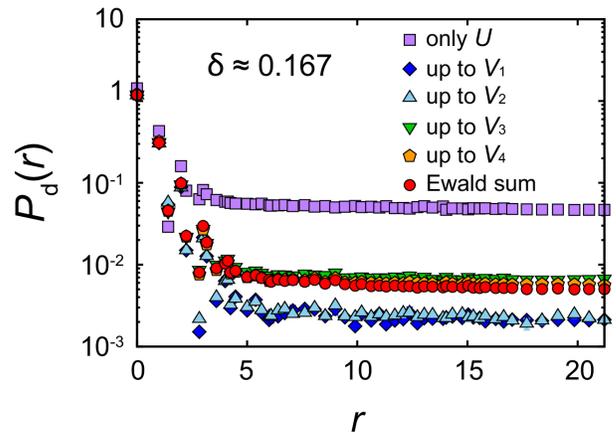}
\caption{(Color online) $P_d(r)$ at $\delta \simeq 0.167$ ($L=30$) for several cases of interaction ranges. In the legends,``only $U$" means that we truncated the off-site interactions and ``up to $V_i$" means that we included them up to $V_n$.}\label{fig:Vdep}\end{center}
\end{figure}

In the previous study on the Hubbard model, controversial results were reported: the nearest-neighbor interaction works destructively for $d$-wave superconductivity in a variational Monte Carlo study\cite{misawa2014}, while a dynamical mean-field (DMFT) study showed the insensitivity of superconductivity to the nearest neighbor repulsion\cite{senechal2013}. Effects of off-site Coulomb interactions beyond the nearest-neighbor pair were mostly neglected in the literature. To clarify the role of the realistic off-site interactions, we have studied the interaction-range dependence of $P_d(r)$ by switching off specific long-ranged parts of $V_{ij}$ from the {\it ab initio} value as shown in Fig. \ref{fig:Vdep}. It indicates that $V_1$ and $V_3$ have particularly strong effects on the superconductivity. Here, $V_n$  stands for the off-site interaction for the $n$-th neighbor pair. $V_1$ and $V_3$ both work in the directions along the Cu-O-Cu bonds. The destruction by $V_1$ is consistent with the result in \cite{misawa2014}. (Note that the short-ranged part of $P_d(r) (r<2)$ are not largely affected by $V_1$. When we consider the mean-field nature of DMFT, which takes into account only the short-ranged correlations by regarding them as mean fields, this insensitivity is also consistent with \cite{senechal2013}. Nevertheless, the true long-range order to be examined in the long-ranged part is severely suppressed by $V_1$.)  The partial recovery by including $V_3$ can be ascribed to the frustrative competition with $V_1$. Eventually, the full {\it ab initio} interactions reduce the superconducting long-range order from the case with $U$ only by nearly one order of magnitude. We note that the result ``up to $V_4$'' is already close to the ``Ewald sum''. This indicates that short-range part of off-site Coulomb repulsions predominantly determines the superconductivity because of the short coherence length (Cooper pairs are formed locally in real space).

A more important effect of off-site interaction is observed in the energy competition between the SC state and stripe states. Without the off-site Coulomb interactions, they are almost degenerate (see Appendix E). Therefore, the off-site Coulomb interactions play a crucial role in energetically stabilizing the SC state against stripe states. Note that the Hubbard model with only the onsite interaction and the nearest-neighbor transfer even more favors the stripe states\cite{zheng2017,darmawan2018}.

\section{\label{sec:chi_c}Connection between charge fluctuations and superconductivity}

\begin{figure}[h]
\begin{center}
  \includegraphics[width=8.5cm]{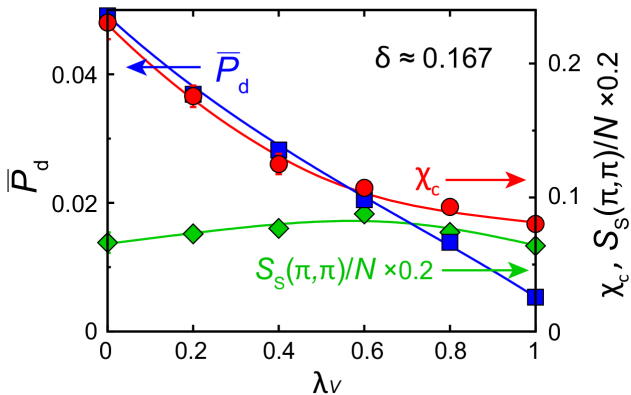}
\caption{(Color online) Comparison among superconductivity, spin structure factor, and uniform charge susceptibility. ${\overline P}_d$(blue squares), $S_s(\pi,\pi)$(green diamonds) and $\chi_c$(red circles) as functions of $\lambda_V$ at $\delta \simeq 0.167$ ($L=30$). Here, we included the off-site Coulomb interactions up to the 4th neighbor one ($V_4$). Note that the long-range tail of the interaction beyond $V_4$ have little effect on the superconductivity [see Fig. \ref{fig:phys_uniform} (b)]. In the presence of the true long-range Coulomb interaction, the uniform charge susceptibility becomes zero, while the realistic metallic screening between different layers can make it short-ranged, and thus $\chi_c$ has finite values in realistic situations. Therefore, we here analyze $\chi_c$ with this finite cutoff to compare its trend as a function of $\chi_c$ with ${\overline P}_d$.}\label{fig:chi_c}
\end{center}
\end{figure}

In the previous studies, the tight connection between the enhancement of superconductivity and that of charge fluctuations was observed in the homogeneous states of the simple Hubbard model\cite{misawa2014}. To examine the relevance of charge and spin fluctuations in the case of realistic Hg-based cuprates, we introduce a single parameter $\lambda_V$ which rescales all the off-site Coulomb interactions $V$ uniformly and thus enables us to monitor the effect on the superconductivity. More precisely, we consider the Hamiltonian where the off-site Coulomb interaction terms ${\cal H}_V$ is replaced by $\lambda_V {\cal H}_V$. In Fig. \ref{fig:chi_c}, we show ${\overline P}_d$, $S_s(\pi,\pi)$, and the uniform charge susceptibility $\chi_c$ as functions of $\lambda_V$ at $L=30$ and $\delta \simeq 0.167$. Here, $\chi_c$ is defined by $d \langle n \rangle/d \mu$ ($\mu$ is the chemical potential) and it was obtained from the calculated $\mu-\delta$ curves (see Appendix F). As seen in this figure, the enhancement of ${\overline P}_d$ is accompanied by that of $\chi_c$ rather than $S_s(\pi,\pi)$ (spin correlation). This shows that charge fluctuations or the resulting effective attraction between carriers is crucial for the enhancement of superconductivity, whereas it also causes the competing inhomogeneity (stripes).

\section{\label{sec:sum}Discussion and summary}

We here discuss the issue on the origin of the condensation energy in more detail. To elucidate whether large interactions induce the crossover to the kinetic-energy driven superconductivity near the {\it ab initio} Hamiltonian, we here introduce a single parameter $\lambda_{\rm int}$ which rescales all the interaction term ${\cal H}_{\rm int}$ uniformly as $\lambda_{\rm int} {\cal H}_{\rm int}$. The $\lambda_{\rm int}$-dependence of the condensation energies is shown in Fig. \ref{fig:e_cond} (b). This shows that $\lambda_{\rm int} >1.8$ $(U/t_1 >13.6)$ is required for $\Delta E_{\rm kin} > 0$, much larger than the {\it ab initio} value $\lambda_{\rm int}$=1. (If we employ the crossover point as the crossing of $\Delta E_{\rm kin}$ and $\Delta E_{\rm int}$, it is even as large as $\lambda_{\rm int}=2.1$). Such a large $U/t_1$  required for the crossover is more or less consistent with the previous VMC studies on the Hubbard model in similar regions of $\delta$\cite{yokoyama2013,tocchio2016}. In the cDMFT and the dynamical cluster approximation (DCA) studies of the $t$-$J$ or Hubbard models\cite{gull2012,fratino2016,haule2007}, aside from the variety of the results not necessarily consistent each other, it was reported that the SC state can become kinetic-energy driven above relatively small values of $U/t_1 \sim 5.5$ at low doping concentration. To realize a kinetic-energy driven SC state for $\delta>0.1$, large $U/t_1 (\simeq 9)$ was reported to be still necessary in the cDMFT study\cite{fratino2016}.  A very large $U$ required for the crossover in the present Hamiltonian in comparison to the cDMFT and DCA may be ascribed partly to the realistic off-site interaction which effectively compensates the contribution from $U$ and another possible origin is the real antiferromagnetic correlation underestimated in the cDMFT and DCA.

In the optical experiments on the cuprates, it was reported that the SC state is driven by a reduction of the kinetic energy in the underdoped region\cite{carbone2006}, which is consistent with the cDMFT studies on the Hubbard model at large $U/t_1$\cite{gull2012,fratino2016} and $t$-$J$ model\cite{haule2007}. However, this looks different from the present result at least at {\it ab initio} parameters. We here discuss the origin of this apparent discrepancy.

The total condensation energy appears to be in the order of 1K commonly in the cuprates as indicated by the specific heat measurement\cite{loram2000} while the kinetic energy gain integrated up to 1.25 eV has the order of 10K\cite{carbone2006}. On the other hand, recent ellipsometer measurement suggests that the Coulomb energy loss contributed from small momentum $|{\bm q}|$ has the energy scale of only 1K\cite{levallois2016}. These somewhat puzzling feature implies that the kinetic energy gain at lower energy below the above cutoff could cancel the loss at higher energy contribution and/or the interaction energy gain/loss could be distributed over a wide $|{\bm q}|$ region. These possibilities are compatible with the present result. First, our result indicates that the Coulomb energy gain must come from the onsite Coulomb interaction part related to the double occupation of two electrons and this local character means that the gain must be distributed more or less uniformly in a wide momentum area beyond the accessible range in Ref. \cite{levallois2016} (see Appendix G). In addition, the main contribution measured in optics to the condensation energy should come from the energy scale of the Mott gap (double occupation energy) and therefore beyond the experimental energy cutoff in Ref. \cite{levallois2016}.  Correspondingly, the kinetic energy loss in the present results may also be distributed in the high-energy range again beyond the optical energy cut off in Ref. \cite{carbone2006}. Although the high energy part is overlapped with the interband transition and is difficult to resolve in experiments for the moment, it is crucial to test the present first-principles result by the accurate high-energy or short-time probe.

In summary, we have studied superconductivity and inhomogeneity in ${\rm Hg Ba}_2 {\rm Cu O}_{4+y}$ by solving an {\it ab initio} low-energy effective Hamiltonian derived before \cite{hirayama2019} with an accurate numerical method. We have found that the charge uniform $d$-wave superconductivity dominates the phase diagram in the ground state in a wide region of doping concentration at $\delta > 0.1$ in agreement with the experimental phase diagram and in contrast to the result of the simple Hubbard model. Furthermore, we found that the off-site Coulomb interactions dramatically reduce the amplitude of superconductivity, while they greatly contribute to lowering the relative energy and stabilizing the superconducting state against the severely competing stripe phases. The driving force of the superconductivity to gain the condensation energy is the onsite interaction energy, where the $d$-wave superconducting state greatly reduces the energy cost of the electron double occupation by the $d$-wave pairing symmetry, where the double occupation is strictly excluded.  This energy gain is represented in the high-energy part of the dynamics involving the doubly occupied sites and is not experimentally accessible so far. It is desired to test this prediction in refined measurements.

On the other hand, the stripe state appears as the ground state in the limited underdoped region around $\delta \simeq 0.1$ and the wavelength of the charge order described by charge 4 and spin 8 lattice constants. These are again consistent with the experiments, but in contrast to the simple Hubbard model. Further low doping region $\delta < 0.1$ is dominated by the antiferromagnetic order as is expected.

All of these show that {\it ab initio} parameters are crucial to reproduce physics of high-$T_c$ cuprates quantitatively. By monitoring the off-site Coulomb interactions beyond the {\it ab initio} values, enhanced charge fluctuations are demonstrated to synchronize with superconductivity.

For deeper and more precise understanding of their physics, studies on the dynamical properties, finite temperature properties and roles of electron-phonon couplings\cite{ohgoe2014,ohgoe2017} are desired in future studies based on the present basic successful understanding. The success in the present Hg-based compound urges more thorough studies on other cuprate and iron-based superconductors in the same first-principles framework to deepen the understanding on the universality and individual character of the Hg compound, which then help designing of better and higher-$T_c$ superconductors.

\section*{ACKNOWLEDGMENTS}

The authors thank T. Tadano and Y. Nomura for useful continuous discussions. We also thank A. S. Darmawan and H.-H. Zhao for providing the extended mVMC code where the fat-tree tensor network is implemented. This work is financially supported by the MEXT HPCI Strategic Programs, and the Creation of New Functional Devices and High-Performance Materials to Support Next Generation Industries (CDMSI). This work was also supported by a Grant-in-Aid for Scientific Research (No. 22104010, No. 22340090, No. 16H06345 and No. 18K13477) from MEXT, Japan. The simulations were partially performed on the K computer provided by the RIKEN Advanced Institute for Computational Science under the HPCI System Research project (the project number hp130007, hp140215, hp150211, hp160201, hp170263 and hp180170). The simulations were also performed on computers at the Supercomputer Center, Institute for Solid State Physics, University of Tokyo.

\appendix

\section{Earliar first-principles studies}
Although most of theoretical studies employ adjustable parameters without any derivations, there exist several efforts to derive parameters of Hubbard-type or $t$-$J$-type models based on first principles (see for example Refs.\cite{munoz2002,furness2018,nilsson2019,sakakibara2019,anisimov2002}). However, except for very few cases, derived effective Hamiltonians were not solved to see whether the solution really reproduces the phase diagram of the cuprates including the superconductivity severely competing with the spin-charge stripe states.

 In Ref. \cite{munoz2002}, $t$-$J$ effective Hamiltonian parameters are derived using quantum chemical analysis of small clusters.  However, they did not solve the derived Hamiltonian and it is not clear whether the simple $t$-$J$ model with only the nearest neighbor interaction captures the experimental phase diagram of the cuprates quantitatively with the severe competitions. In fact, we have shown the importance of the off-site interactions to understand the severe competition between the superconductivity and stripes, while such an issue is neglected by ignoring the possible competition with charge inhomogeneities.

In Ref. \cite{furness2018}, by improving the density functional theory, the antiferromagnetic insulating properties for ${\rm La}_2{\rm Cu O}_4$ were reproduced, while its insulating gap totally relies on the antiferromagnetic order and the Mott insulating nature is missing. The central question of the superconductivity was not studied anyway.

In Refs. \cite{nilsson2019} and \cite{sakakibara2019}, effective Hamiltonians for a few cuprate compounds were derived.  The Hamiltonian parameters have an overall consistency between the present Hamiltonian and that in Ref. \cite{nilsson2019}, while the onsite interaction parameter derived in Ref. \cite{sakakibara2019} is substantially smaller than our value. The main reason is that they did not use a proper disentanglement procedure for the entangled bands employed in Ref. \cite{hirayama2019}.  Another origin of the discrepancy is that Ref. \cite{sakakibara2019} derived the Hamiltonian so as to ignore the offsite interaction. In both of Refs. \cite{nilsson2019} and \cite{sakakibara2019}, severe competitions between the superconducting and stripe states and their carrier concentration dependence were not studied.

Reference \cite{anisimov2002} derived the effective $t$-$J$ Hamiltonian by using the constrained LDA method and solved it by the variational Monte Carlo method. In the derivation of the effective $t$-$J$ Hamiltonian, various refined treatments developed recently including the cGW method employed in the present study were not taken into account. Ref. \cite{anisimov2002} employing the strong-coupling limit reproduced some feature of superconductivity, while various important aspects such as the role of off-site Coulomb interaction, which must be much larger than $J$ and could easily destroy the superconductivity, and the severe competition with the static stripe phase were not seriously examined and the importance of the charge fluctuation was not considered. Recent more refined studies on a $t$-$J$ model using the tensor network proposed the coexistence of superconductivity with stripes, although there remains uncertainty in its extrapolation with respect to the inverse tensor dimension\cite{corboz2014}.

The present study overcomes many of the limitations in the previous studies, in terms of the level of reliability and quantitative accuracy of the effective Hamiltonian as well as the accuracy of the solver as clarified in Refs. \cite{darmawan2018} and \cite{hirayama2019}.

\section{Ewald summation}

\begin{figure}[H]
\begin{center}
  \includegraphics[width=8cm]{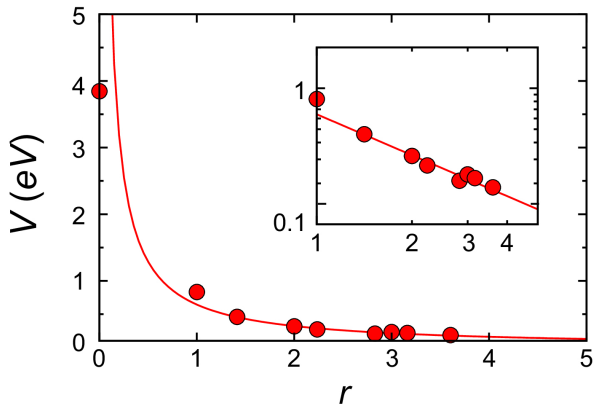}
\caption{(Color online) $r$-dependence of the screened Coulomb interactions for the {\it ab initio} single-band effective Hamiltonian of the Hg-based cuprate. As the unit of the distance, we use the distance between the nearest-neighbor Cu atoms in the ${\rm Cu O}_2$ plane. The inset shows the logarithmic plot. The red curve (line in the inset) is obtained by the $1/r$ fitting in the long-ranged part.}\label{fig:vr}\end{center}
\end{figure}

Here we briefly describe how we treated the long-range part of the screened Coulomb interaction $V$ in our Hamiltonian. In Fig. \ref{fig:vr}, we show the {\it ab initio} screened Coulomb interaction $V$ as a function of the relative distance $r$. As seen in the logarithmic plot of the inset, the long-range part decays as $\sim 1/r$ for large $r$ and we determined the coefficient by fitting. Then, in our simulations of finite systems, we employed the Ewald summation to include the long-range part of the screened Coulomb interaction $V$ accurately.

\section{Connection to the Hubbard model}

Since the condensation energy for the Hg-based cuprate is as small as 0.1 meV, reproducing its value with high accuracy is beyond the ability of the present numerical approach, because our errors are typically 1 or a few meV after the variance extrapolation. However, the observed energies, which is close between the SC state and the normal metal do not contradict experiments. Still, it is instructive to show that the observed SC state is adiabatically connected to the case where one can clearly establish the superconducting ground state with resolved positive condensation energy within the numerical accuracy. This is the case of the simple Hubbard model (with only $t_1$ and $U$) at a specific hole density and $U/t_1$ . In the Hubbard model, the SC state has been more clearly shown to be the ground state around $\delta \simeq 0.2$ in a recent study\cite{darmawan2018}.

\begin{figure}[H]
\begin{center}
  \includegraphics[width=7cm]{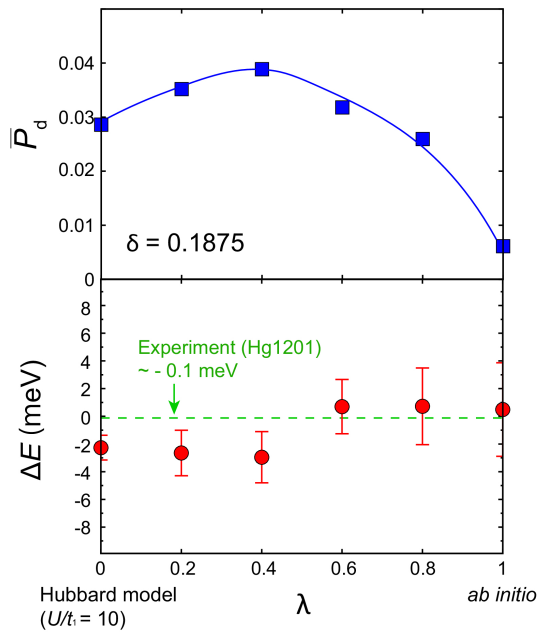}
\caption{(Color online) $\overline{P}_d$ and $\Delta E$ as functions of $\lambda$ at $\delta=0.1875$ ($L=24$). $\Delta E$ is obtained after the variance extrapolation of energies. The dashed line in the lower panel represents the experimental value ($\Delta E = -E_{\rm cond}$)\cite{billon1997,kirtley1998}.}\label{fig:connectToHB}\end{center}
\end{figure}

To connect the Hubbard model at $U/t_1 =10$ to the {\it ab initio} Hamiltonian, we introduce a single parameter $\lambda$ which uniformly rescales all the parameter difference between the two Hamiltonians. We define $\lambda$ such that $\lambda=0$ and 1 correspond to the Hubbard model and the {\it ab initio} Hamiltonian, respectively, and $\lambda$ linearly interpolates these two limits. In Fig. \ref{fig:connectToHB}, we show $\overline{P}_d$ and $\Delta E$ as functions of $\lambda$ at $\delta=0.1875$ ($L=24$). From $\overline{P}_d$, we see that the superconducting state at $\lambda=0$ smoothly connects to $\lambda=1$. In addition, $\Delta E$ is positive on the  $\lambda=0$ side. At $\lambda=0$, the competition with stripe states was also studied in Ref. \cite{darmawan2018}, and it has been shown that the energies of homogeneous states are lower than those of stripe states around $\delta \simeq 0.2$ including $\delta=0.1875$. Therefore, we conclude that the SC state is the ground state at $\lambda=0$.

\section{Size dependence of stripe orders }

\begin{figure}[H]
\begin{center}
  \includegraphics[width=7cm]{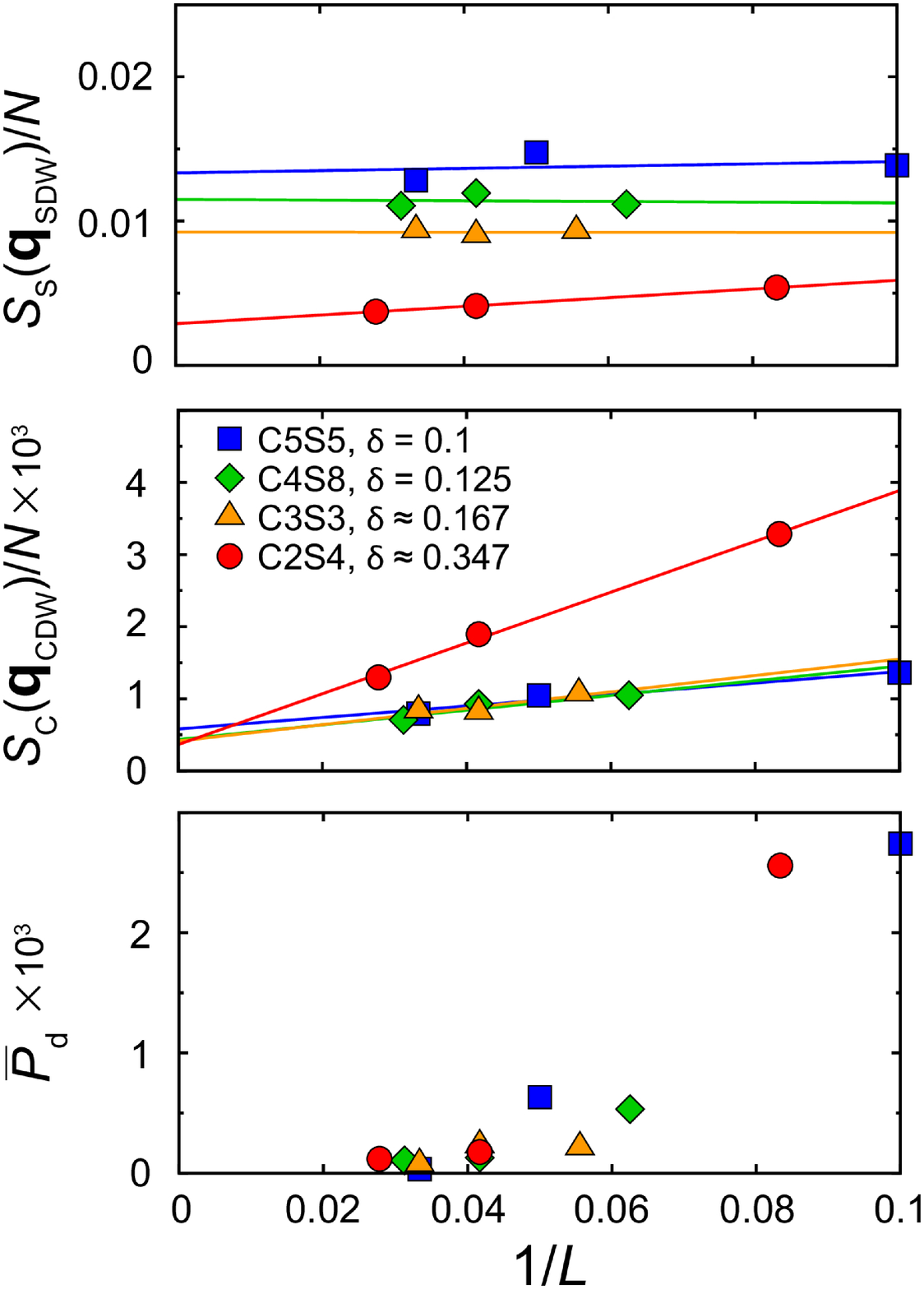}
\caption{(Color online) Size dependence of $S_s({\bm q}_{\rm SDW})/N$, $S_c({\bm q}_{\rm CDW})/N$ and ${\overline P}_d$ for each stripe. We performed linear extrapolations of $S_s({\bm q}_{\rm SDW})/N$ and $S_c({\bm q}_{\rm CDW})/N$ to the thermodynamic limit, which are shown as lines.}\label{fig:stripes_sizedep}\end{center}
\end{figure}

In Fig. \ref{fig:stripes} (a) of the main text, we showed $S_s({\bm q}_{\rm SDW})/N$ and $S_c({\bm q}_{\rm CDW})/N$ of stripe states. Here, we show the size dependence of them. In Fig. \ref{fig:stripes_sizedep}, we plot the structure factors as functions of $1/L$. We also include ${\overline P}_d$ to show its superconductivity. The linear extrapolations to the thermodynamic limit indicate that both the spin and charge orders are long ranged. On the other hand, ${\overline P}_d$ is strongly suppressed for larger systems, demonstrating the nature of competition between superconductivity and stripes.

\section{Energy competition without off-site Coulomb interactions }

\begin{figure}[h]
\begin{center}
  \includegraphics[width=8cm]{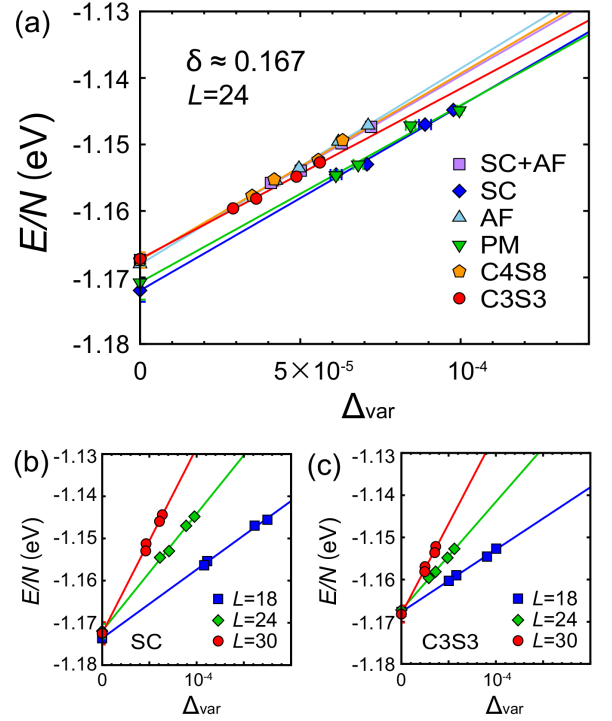}
\caption{(Color online) Extrapolation of energies to zero variance limit. (a) Extrapolations of energies per site
$E/N$ of different states to the variance $\Delta_{\rm var}$=0 ($L=24$ and $\delta \simeq 0.167$). Here, $\Delta_{\rm var}$ is defined by $\Delta_{\rm var} = (\langle H^2 \rangle - \langle H \rangle^2)/\langle H \rangle^2$. In the legend, ``SC+AF'' and ``PM'' represent a coexisting state of SC and AF, and a paramagnetic metal, respectively. Each state has four energies obtained by different methods: 1) mVMC method, 2) mVMC + fat-tree tensor network (FTTN) method, 3) mVMC + 1st Lanczos method, and 4) mVMC + FTTN + 1st Lanczos method. The energies are decreasing in this order. For FTTN, we used the bond dimension $D=2$. (b) and (c) Variance extrapolations of SC states and C3S3 states for different system sizes, respectively. $\Delta_{\rm var}$ becomes smaller for larger system sizes because it should scale as $1/N$. The extrapolated energies for different system sizes agree with each other within error bars.}\label{fig:var_extrapolation}\end{center}
\end{figure}

\begin{figure}[H]
\begin{center}
  \includegraphics[width=8cm]{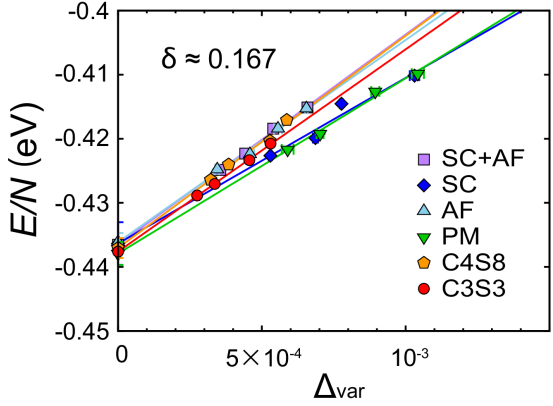}
\caption{(Color online) Variance extrapolation of energies of different states for the Hamiltonian without the off-site Coulomb interactions ($L$=24).}\label{fig:ene_uonly}\end{center}
\end{figure}

To understand the role of off-site Coulomb interactions on energy competitions among different states, we here present the results of the energy competition without off-site Coulomb interactions. We first show the variance extrapolation of energies for the {\it ab initio} Hamiltonian in Fig. \ref{fig:var_extrapolation}. Figure \ref{fig:ene_uonly} shows the variance extrapolation of energies where the only difference is that we here switched off the off-site Coulomb interactions. As a result, the extrapolated energies become very close, and thus the SC state is more severely competing with stripe states. This shows that the off-site Coulomb interactions play a crucial role of energetically stabilizing the SC state against stripe states.

\section{Uniform charge susceptibility}
\begin{figure}[H]
\begin{center}
  \includegraphics[width=8cm]{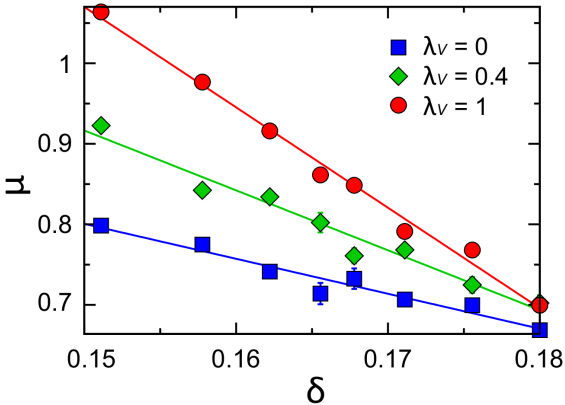}
\caption{(Color online) $\mu - \delta$ curves for $\lambda_V$=0, 0.4 and 1 ($L$=30). Unimportant constant terms are shifted for clarity. The results are obtained by the mVMC method without variance extrapolations.}\label{fig:mu}\end{center}
\end{figure}

Here we explain how we obtained the uniform charge susceptibilities $\chi_c$ shown in Fig. \ref{fig:chi_c} of the main text.
It is defined by $\chi_c=d\langle n \rangle/d\mu$, where $\mu$ is the chemical potential.
To obtain it, we first calculated total energies $E$ at two different electron numbers $N_e$ and $N'_e$ which are close to each other.
Then, we evaluated $\mu$ at the middle filling $\overline{N}_e=(N_e + N'_e)/2$ as $\mu(\overline{N}_e) = [ E(N_e) -E(N'_e) ]/(N_e - N'_e)$.
After we obtain the $\mu - \delta$ curve, we performed a linear fitting near $\delta \simeq 0.167$ to estimate the slope.
Since $\chi_c^{-1} = -d\mu/d\delta$, we can finally obtain $\chi_c$ as the inverse of the negative slope.
Figure \ref{fig:mu} shows the $\mu - \delta$ curves and the results of fittings for $\lambda_V=$0, 0.4, and 1.

\section{Momentum resolved condensation energy in interaction- vs. kinetic-energy parts}

\begin{figure}[H]
\begin{center}
  \includegraphics[width=8.5cm]{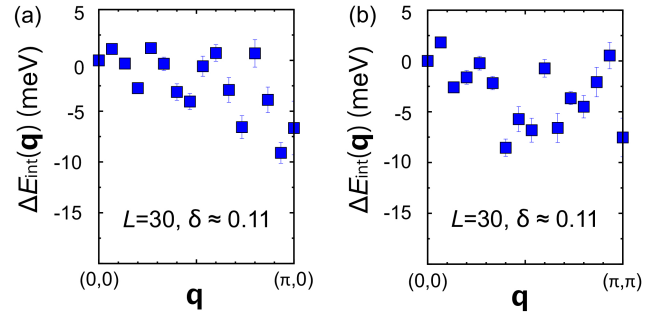}
\caption{(Color online) Super-Normal energy difference $\Delta E (=-E_{\rm cond})$ coming from the momentum-resolved Coulomb interaction energy $E_{\rm int}({\bm q})$. (a) Dependence along the symmetry line from $(0,0)$ to $(\pi,0)$. (b) Dependence along the symmetry line from $(0,0)$ to $(\pi,\pi)$.  }\label{fig:eint_cond_q}\end{center}
\end{figure}

In recent temperature-dependent ellipsometry measurements on Bi-based cuprates\cite{loram2000}, the partial Coulomb energy for the wave vector ${\bm q} \sim 0$ was measured. They reported that the Coulomb energy gain around the wave vector ${\bm q} \sim 0$ to stabilize superconductivity is comparable to the total condensation energy ~1K reported in the specific heat measurement with a similar tendency for the doping concentration dependence\cite{levallois2016}. However, the whole kinetic energy loss or gain as compared to the normal state has the scale of 10K\cite{carbone2006}, one order of magnitude larger than the interaction energy gain/loss coming from the small ${\bm q}$ region as inferred from the optical conductivity measurement. This implies that there is much larger energy scale distributed in the large ${\bm q}$ region (including ${\bm q} \sim (\pi,\pi)$ region) of the interaction energy to compensate the kinetic energy gain/loss and to stabilize the experimental superconducting state with the positive condensation energy of the order 1K. To gain insight from the theoretical analysis of the {\it ab initio} Hamiltonian, we calculated the ${\bm q}$-resolved Coulomb interaction energy $E_{\rm int} ({\bm q})$ and its energy difference $\Delta E_{\rm int}({\bm q}) = E_{\rm int, SC}({\bm q})/N - E_{\rm int, Normal}({\bm q})/N$. Here, we define $E_{\rm int} ({\bm q})$ from $E_{\rm int} = \frac{1}{2N} \sum_{\bm q} E_{\rm int}({\bm q})$ and $E_{\rm int} ({\bm q}) = V ({\bm q}) \langle n_{\bm q} n_{-{\bm q}} \rangle$. $V ({\bm q})$ is the Fourier transformation of the sum of all the screened Coulomb interaction after the Ewald summation. The results along two symmetric directions in the Brillouin zone are shown in Figs. \ref{fig:eint_cond_q} (a) and (b). Although the resolution of the available data is not sufficient enough in this tiny energy scale, we find a trend of large energy gain in a wide ${\bm q}$ region, which is consistent with the above experimental indications and the intuition from the local energy gain addressed in the main text as the main energy gain in the onsite interaction part associated with the double occupation energy.

\bibliography{reference}

\end{document}